# An Effort to Measure Customer Relationship Performance in Indonesia's Fintech Industry

[1]Alisya Putri Rabbani, [2]Andry Alamsyah, [3]Sri Widiyanesti
[1,2,3]School of Economic and Business, Telkom University, Bandung, Indonesia
[1]alisyaprabbani@student.telkomuniversity.ac.id, [2]andrya@telkomuniversity.ac.id, [3]widiyanesti@telkomuniversity.ac.id

*Abstract*— The availability of social media simplifies the companies-customers relationship. An effort to engage customers in conversation networks using social media is called Social Customer Relationship Management (SCRM). Social Network Analysis helps to understand network characteristics and how active the conversation network on social media. Calculating its network properties are beneficial for measuring customer relationship performance. Financial Technology, a new emerging industry that provides digital-based financial services utilize social media to interact with its customers. Measuring SCRM performance is needed in order to stay competitive among others. Therefore, we aim to explore SCRM performance of Indonesia's Fintech company. In terms of discovering the market majority thought in conversation networks, we perform sentiment analysis by classifying into positive and negative opinion. As case studies, we investigate Twitter conversation about *GoPay, OVO, Dana*, and *LinkAja* during the observation period from 1st October until 1st November 2019. The result of this research is beneficial for business intelligence purpose especially on managing relationships with customers.

*Keywords—Social Network Analysis, Customer Relationship Management, Sentiment Analysis, Financial Technology, Business Intelligent, Social Media*

## I. Introduction

Social media users have increased in numbers years to years. In Indonesia, there are 150 million people recorded as social media active users in 2019 [1]. People access social media to connect each other, search for information, share their interest or experience, and express their opinion about the product or service they have used. On the other hand, companies from various sectors use social media for their marketing activities. Social media allows companies to manage online conversation networks with both customers and the market. Moreover, social media also leads the companies to a new approach for engaging more customers by seeing them as a community instead of an individual. This activity is called social customer relationship management (SCRM) [2].

As one of the most popular social media, Twitter is powerful in conversation network formation. Besides, Twitter is capable of capturing users' emotions in a short-post. By providing some simple features such as tweet, reply, share, and like, Twitter succeeded in obtaining the large-scale database having 200 million users who post 400 million tweets in a day [3].

Fintech, an innovative technology-enabled financial service is fast becoming a global phenomenon [4]. In Indonesia, fintech companies are increasingly emerging as more people familiar with it. Social media play a role in this rapid growth. *GoPay, OVO, Dana*, and *LinkAja* are the leading on fintech in Indonesia [5]. All of these companies take advantage of social media to share information about their services, respond to complaints, interact, and engage with their customers.

Social Network Analysis (SNA) helps to understand network characteristics since it provides several metrics and measurement [6]. Moreover, SNA is able to visualize networks easier based on graph theory [7]. The activities of online social network conversations enhance since its numbers of internet and social media users are increasing. Those help the business to extract valuable information to improve business performance and the ability to achieve customer engagement [8]. Online conversation networks consist of users' opinion which really important for companies because it will affect brand image. Sentiment analysis helps to capture users' majority thought by classifying into positive and negative opinion.

In this research, we aim to measure social network performance and identify market opinion about Indonesia's fintech on Twitter. We perform SNA by calculating its network properties and compare the results of each fintech i.e. *GoPay, OVO, Dana*, and *LinkAja*. We also utilize sentiment analysis to discover which fintech has the most positive thoughts. Several previous researches have been used to support this research, such as the use of SNA and sentiment analysis to measure SCRM performance and the empirical links between network characteristics and social engagement [8][9]. The results are beneficial for each company to understand both of their customers and competitors deeper and better. Moreover, this insight will help companies evaluate and make the decision for their future SCRM strategies.

## II. Theoretical Background

### A. Social Customer Relationship Management

Customer relationship management (CRM) is defined as a comprehensive strategy and process of acquiring, retaining, and partnering with selective customers to create superior value for the company and the customer that involves the integration of marketing, sales, customer service, and the supply-chain functions of the organisation to achieve greater efficiencies and effectiveness in delivering customer value [10]. The availability of social media make companies perform CRM effort much easier, this is called social customer relationship management (SCRM).

The relationship between SNA and SCRM is that SNA extend and enhance the capabilities of SCRM with new methodological way to engage customers and manage the conversations [8]. Therefore, SNA could be considered as a method to measure SCRM effort because give businesses a way to manage and measure their performance in engaging users.

*B. Social Network Analysis*

Social network analysis (SNA) is a methodology for studying the connections and behavior of individuals within social groups [11]. Consist of nodes that represent actor and edges that represent relations between actors, these network model provide several metrics based on theory graph characteristic. It is important to do SNA if one wants to understand the structure of the network so as to gain insights about how the network 'works' and make decisions upon it by either examining node/link characteristics or by looking metrics at the whole network cohesion [12]. SNA can be applied to a wide range of business problem such as strategy, sales and marketing, human resource, team-building, also knowledge management and collaboration [13].

In this research, we calculate network properties with SNA approach to measure social customer performance. This network properties have been used in previous research to support the SCRM Network [8][9]. Table I shows definition of each network properties.

TABLE I. NETWORK PROPERTIES DESCRIPTION

| No. | Description of Network Properties | |
|---|---|---|
| | Network Property | Description |
| 1 | Size | Number of nodes in network |
| 2 | Density | The fraction of number edges in network to the maximum edges possible |
| 3 | Modularity | The fraction of edges within communities minus the expected value of that fraction if the position of the edges are randomized |
| 4 | Diameter | The largest distance recorded between any pair of nodes |
| 5 | Average Path Length | The average distance between all pairs of nodes in the network |
| 6 | Average Degree | The average number of links a node has to the other node |
| 7 | Reachability | The fraction of node pairs that are connected |
| 8 | Connected Component | Measure of maximal subset of nodes such that each node is reachable by some paths from each other |

Source: [9]

*C. Sentiment Analysis*

Sentiment analysis, also called opinion mining is the field of study that analyzes people's opinions, sentiments, evaluations, appraisals, attitudes, and emotions towards entities such as products, services, organizations, individuals, issues, events, topics, and their attributes [14]. Automated sentiment analysis is mandatory because the average human reader has difficulty in identifying relevant sites, extracting and summarizing the opinions in them. Several classification methods are suitable to perform sentiment analysis, such as Support Vector Machine, Naïve Bayes, Decision Tree, and Nearest Neighbors. In this research we use Naïve Bayes classification method because it is more efficient and more accurate compared to other classifiers. Previous research said that Naïve Bayes is more computationally efficient because it allows each attribute to contribute towards the final decision equally and independently, has high-level accuracy and support large data processing [15].

*D. Social Media*

Social media is an online platform which people use to build social networks or social relations with other people who share similar personal or career interests, activities, backgrounds or real-life connections [16]. Social media is beneficial for individual, professional, and business. On an individual level, social media allow people to communicate with each other, gain knowledge of new things, develop interests, and be entertained. On a professional level, social media can be used to expand or broaden knowledge in a particular field and build our professional network by connecting with other professionals in same industry. At the business level, social media allows to have a conversation with audience, gain customer feedback, and elevate brand.

Microblogging is one of social media form with high potency because of its direct impact on the formation of electronic word of mouth (EWOM) and also allow the user to share their interest and express attitudes that they are willing to share with others in short posts, either it's positive or negative [17]. Twitter is the most popular microblogging with the large-scale database having 200 million users who post 400 million tweets in a day [3].

## III. METHODOLOGY

Our methodology research consists of four stages. The first stage collect data from Twitter and preprocess data. Second, construct network model, followed by measurement of the 8 network properties. Third, perform sentiment analysis to discover the percentage of each sentiment. The last one, compare both network properties and sentiment results to find out which fintech has the best SCRM performance. The research workflow is shown in Figure 1.

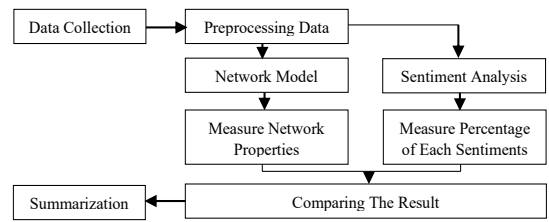

Fig. 1 Research Workflow

*A. Data Collection and Preprocessing Data*

In this process, we crawl tweets data using Twitter Data Streaming API. We define the keyword related to the topics before collecting tweets, there are *'gopayindonesia'*, *'ovo_id'*, *'danawallet'*, and *'linkaja'*. The period of data collection is from 1st October 2019 until 1st November 2019. And then, we complete data preprocessing to support the methods applied in this research and get relevant data to the research objectives. For SNA method, we preprocess data by filtering data attributes that are required to define the nodes and edges. For sentiment analysis method, we manage three steps to get accurate results: tokenizing to separate sentences into a word, word filtering to remove common words, and stemming to refer the word to its original form.

## B. Social Network Analysis

In this process, we construct the network models for each of fintech companies to see the pattern of interactions between users that share their opinion about *GoPay*, *OVO*, *Dana*, and *LinkAja*. And then, we calculate the network properties with *Python*. There are size, density, modularity, diameter, average path length, average degree, reachability, and connected component. Table 2 shows the formula of each network properties.

TABLE II. NETWORK PROPERTIES FORMULA

| No. | Network Properties | Formula |
|---|---|---|
| 1. | Size | Number of nodes in network [9]. |
| 2. | Density | $D = \dfrac{L}{g(g-1)}$ <br> $D$ is a density value, $L$ is the actual relationship, $g$ is network size [18]. |
| 3. | Modularity | $Q = \dfrac{1}{2m}\sum_{ij}\left[A_{ij} - \dfrac{k_i k_j}{2m}\right]\delta_{si.sj}$ <br> $Q$ is modularity value, $m$ is number of edges, $A_{ij}$ is actual number of edges between $i$ and $j$, $k_i k_j$ is the expected number, $\delta_{si.sj}$ is kroneckel delta [19]. |
| 4. | Diameter | $d^{max} = (i, j)$ <br> $d^{max}$ is diameter value, i and j are nodes in the network [6] |
| 5. | Average Path Length | $<d> = \dfrac{1}{N(N-1)}\sum_{i.j=i.N} d_{ij}$ <br> $<d>$ is average path length value, $N$ is the total number of node, $d_{i,j}$ is the closeness between node $i$ and $j$ [6]. |
| 6. | Average Degree | $\langle k^{in}\rangle = \dfrac{1}{N}\sum_{i=1}^{N} k_1^{in} = \langle k^{out}\rangle = \dfrac{1}{N}\sum_{i=1}^{N} k_1^{out} = \dfrac{L}{N}$ <br> $k^{in}$ is incoming degree, $k^{out}$ is outcoming degree, $N$ is number of node, $L$ is number of link. [6] |
| 7. | Reachability | $Rch = \dfrac{m}{N^2/2}$ <br> $Rch$ is reachability value, $m$ is number of connected pairs, $N$ is number of nodes [20]. |
| 8. | Connected Component | Subgraph where node of the subgraph is not reachable from another node of the graph. [21] |

Source: [9], [18], [19], [6], [20], [21]

## C. Sentiment Analysis

Here is the detail of performing sentiment analysis with Naïve Bayes classification method. First, we separate tweets data into the data train and data test. Then, we label every tweets by adding sentiment attribute into the data train with two labels, positive and negative. After all the data train labeled, we use machine learning tools for labeling data test automatically. This process also discovers classification performance with several performance evaluation parameter such as precision, recall, F-measure, accuracy, and kappa score. Table 3 shows an example of how we classify the opinion into positive and negative.

TABLE III. SENTIMENT SAMPLE

| No. | Sentiment Sample | |
|---|---|---|
| | Tweets | Sentiment |
| 1. | @linkaja PakeLinkAja bikin aku mudah dalam bertransaksi pembayaran kapan saja dan dimana saja | Positive |
| 2. | @newsplatter Sebenernya 2 aja udah cukup (OVO n GoPay) tp sekarang punya 3, terpaksa banget soalnya bayar e-ticket harus pake linkaja | Negative |

## IV. RESULT AND ANALYSIS

The results of network properties measurement are shown in Table IV. The visualization of *GoPay, OVO, Dana,* and *LinkAja* networks is in Figure 2. The result of sentiment analysis are in Table V. The result of Naïve Bayes text classification performance are in Tabel VI.

TABLE IV. NETWORK PROPERTIES COMPARISON

| No. | Network Properties | GoPay | OVO | Dana | LinkAja |
|---|---|---|---|---|---|
| 1. | Size | 3,478 | 6,495 | 5,682 | 4,531 |
| 2. | Edges | 3,160 | 5,749 | 5,711 | 4,772 |
| 3. | Density | 0.00049 | 0.00025 | 0.00033 | 0.00044 |
| 4. | Modularity | 0.612 | 0.385 | 0.503 | 0.381 |
| 5. | Diameter | 15 | 12 | 13 | 11 |
| 6. | Average Path Length | 3.451 | 2.717 | 3.264 | 3.079 |
| 7. | Average Degree | 1.817 | 1.770 | 2.01 | 2.106 |
| 8. | Reachability | 0.0005 | 0.0002 | 0.0003 | 0.0005 |
| 9. | Connected Component | 586 | 1.324 | 933 | 370 |

Based on the network properties comparison results on Table IV, *OVO* performs best in the number of nodes and edges. The high number of nodes and edges means more users and more interaction between users in the conversation network, thus generate new connections and new opportunities. Besides, *OVO* also performs best in average path length. It has lower value among others which means the chance to precipitate the information dissemination process is increases.

*GoPay* has the highest value of density and modularity. The higher the value of density means the more connections appear in every user in the network. The high number of modularity values means more support in the formation group.

*LinkAja* conversation network performs best in diameter, average degree, and connected component. The lower diameter the better because it means only needed fewer paths to spread information between two users with the longest distance. The higher average degree value means more users get information directly. The lower connected component value the better because it means that the users do not divide too much into small groups [9].

Based on reachability values, both *GoPay* and *LinkAja* have higher numbers, which means their networks manage the information dissemination simpler. The comparison results above point out that the *LinkAja* conversation network performs better by winning 4 out of 9 network properties.

Moreover, we construct the network model of each fintech. The colors of each fintech represent edge or the relationship between nodes, the nodes represent users involved in the conversation network. The network models are shown in Figure 2.

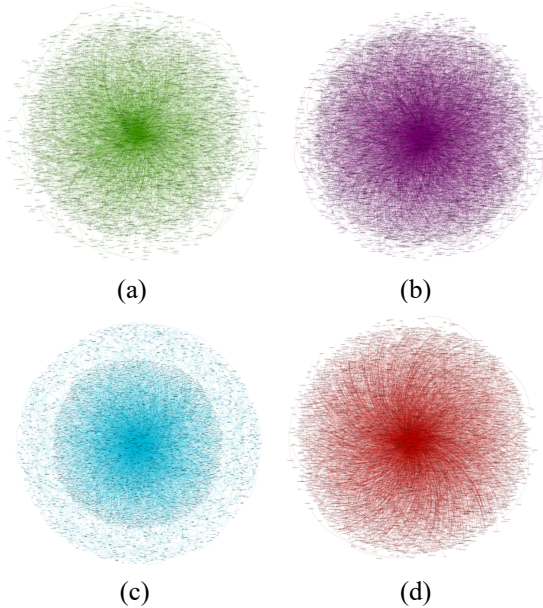

Fig. 2 Network visualization of (a). *GoPay*, (b). *OVO*, (c). *Dana*, and (d). *LinkAja*

TABLE V. SENTIMENT PERCENTAGE COMPARISON

| No. | Sentiment Percentage Comparison | | | | |
|---|---|---|---|---|---|
| | Sentiment | *GoPay* | *OVO* | *Dana* | *LinkAja* |
| 1. | Positif | 36.00% | 47.37% | 53.82% | 72.57% |
| | | 509 tweets | 720 tweets | 818 tweets | 1,119 tweets |
| 2. | Negatif | 64.00% | 52.63% | 46.18% | 27.43% |
| | | 905 tweets | 801 tweets | 702 tweets | 423 tweets |

Table 5 shows the comparison results of sentiment analysis. As we can see, *LinkAja* has the highest numbers of positive sentiment opinion. Followed by *Dana*, with more than a half positive sentiments from total tweets data. *OVO* and *GoPay* have low positive sentiment results compare to the negative ones.

These results discover that both *LinkAja* and *Dana* have been succeeded in creating a positive image for their customers. Otherwise, based on the low results of positive sentiment both *OVO* and *GoPay* must pay attention more to their image in the online conversation network.

Table 6 describes Naïve Bayes text classification performances. The performance evaluation parameter such as precision, recall, F-measure, accuracy, and kappa shows good results. The kappa value of each fintech varies between 0.5 until 0.7 which predicated a moderate agreement and substantial agreement of its classification model, while precision and recall value reach between 75-85%, with F-measure between 75-85% as harmonization of precision and recall value indicating the classification process is going well. Therefore, the Naïve Bayes classification method is suitable to be applied in sentiment analysis for the four fintech data tweets.

TABLE VI. NAÏVE BAYES TEXT CLASSIFICATION PERFORMANCE

| No. | Naïve Bayes Text Classification Performance | | | | |
|---|---|---|---|---|---|
| | | *GoPay* | *OVO* | *Dana* | *LinkAja* |
| 1 | Precision | 83.48% | 78.62% | 77.29% | 81.57% |
| 2 | Recall | 84.79% | 80.91% | 77.74% | 76.51% |
| 3 | F-Measure | 81.43% | 79.75% | 77.51% | 78.96% |
| 4 | Accuracy | 84.55% | 79.03% | 76.06% | 83.56% |
| 5 | Kappa | 0.678 | 0.577 | 0.529 | 0.566 |

## V. CONCLUSION

We have succeeded in performing SNA to measure customer relationship performance. Based on network properties comparison, *LinkAja* conversation network performs better. It makes *LinkAja* have the best SCRM effort than the other although the result does not show a significant difference between four fintech. Along with network performance results, *LinkAja* success in creating a positives image based on sentiment analysis results. *Dana* also succeeds in creating a positives image for their customers. Otherwise, both *GoPay* and *OVO* must pay attention more to their image in the online conversation network.

These results are insightful for business. Companies can evaluate their performance based on network properties, e.g. companies with low value on density and modularity need to create more campaigns or competition related to certain topics to attract more audiences joining the conversation network. Moreover, companies with a low average degree value can collaborate with influential people to create more brand awareness in the conversation network. Companies can also evaluate their performance based on the sentiment analysis result, e.g. companies with low positive sentiments need to do further investigation to their services, being more responsive and friendly to customers, etc.

Here are suggestions for further research: 1) extend the data collection period to enrich the data so that the information generated is more accurate, 2) Identifying SNA metrics such as centrality, betweenness, closeness, and others for exploring further community detection, 3) Using text classification methods other than naïve Bayes in sentiment analysis to explore, and 4) Conduct deeper research to find out from the positive and negative sentiments that are generated, what topics are most discussed using appropriate methods such as topic modeling or other methods.